\documentstyle[twoside,fleqn,espcrc2]{article}
\def\beq{\begin{equation}}
\def\eeq{\end{equation}}
\def\bea{\begin{array}}
\def\eea{\end{array}}
\def\beqa{\begin{eqnarray}}
\def\eeqa{\end{eqnarray}}

\def\u1{{U(1)}}
\def\su2{{SU(2}}
\newcommand{\re}{\relax{\rm I\kern-.18em R}}

\newcommand{\AmS}{{\protect\the\textfont2
  A\kern-.1667em\lower.5ex\hbox{M}\kern-.125emS}}

\title{Hadron Masses and Quark Condensate from Overlap Fermions}
\author{
K. F. Liu
\address{National Center for Theoretical Sciences,
P.O. Box 2-131, Hsinchu, Taiwan}{}
\address{Department of Physics and Astronomy, 
University of Kentucky, Lexington, KY 40506, USA} 
\thanks{Talk presented at Lattice '99, Pisa, Italy},
S. J. Dong$^{\rm \;b}{}$, 
F. X. Lee
\address{Department of Physics, George Washington University, 
Washington, DC 20052, USA}{}
\address{Jefferson Lab, 12000 Jefferson Avenue, Newport News, VA 23606, USA},
and J. B. Zhang$^{\rm \;b\;c}$   
\address{Zhejiang Institute of Modern Physics, Zhejiang University, 
Hangzhou 310027, P.R. China} }
\begin{document}
\begin{abstract}
We present results on hadron masses and quark condensate from Neuberger's 
overlap fermion. The scaling and chiral properties and finite volume 
effects from this new Dirac operator are studied. We find that the
generalized Gell-Mann-Oakes-Renner relation is well satisfied down to
the physical u and d quark mass range. We find that in the range of
the lattice spacing we consider, the $\pi$ and $\rho$ masses at a fixed
$m_{\pi}/m_{\rho}$ ratio have weak $O(a^2)$ dependence.
% on $6^3 \times 12, 8^3 \times 16$, and $10^3 \times 20$ lattices 
%with $\beta = 5.7, 5.85$ and $6.0$.  
\end{abstract}
\maketitle

   The recent advance in chiral fermion formulation which satisfies 
Ginsparg-Wilson relation has a great promise
in implementing chiral fermion for lattice QCD at finite lattice spacing.
It is shown to have exact chiral symmetry on the lattice~\cite{lcz98}  
and it has no order $a$ artifacts~\cite{nie99}. Neuberger's Dirac operator
~\cite {neu98} derived from the overlap formalism has a compact form in four
dimensions which involves a matrix sign function
\begin{equation}  \label{neu}
D = {1\over 2}\left[1 + \mu + (1 - \mu) \gamma_5 \epsilon (H)\right].
\end{equation}

In this talk, I present some preliminary results from 
our numerical implementation of the Neuberger fermion.
We adopt the optimal rational approximation of the matrix sign 
function~\cite{ehn99a} with 12 terms in the polynomials. The smallest
10 to 20 eigenvalues of $H^2$ are projected out for exact evaluation
of the sign function for these eigenstates. We use multi-mass conjugate 
gradient as the matrix solver for both the inner and outer loops. With
residuals at $10^{-7}$, the inner loop takes $\sim 200$ iterations and
the outer loop takes $\sim 100$ iterations. We check the
unitarity of the matrix $V = \gamma_5 \epsilon (H)$. For $Vx = b$, we find
$|x^{\dagger}x - b^{\dagger}b| \sim 10^{-9}$. Even for topological
sectors with $Q \neq 0$, we find the critical slowing down is much milder
than that of the Wilson fermion and there are no exceptional configurations.
The critical slowing down sets in quite abruptly after $\mu a = 0.003$ which
is already at the physical u and d masses. 

\begin{figure}[ht]
\includegraphics{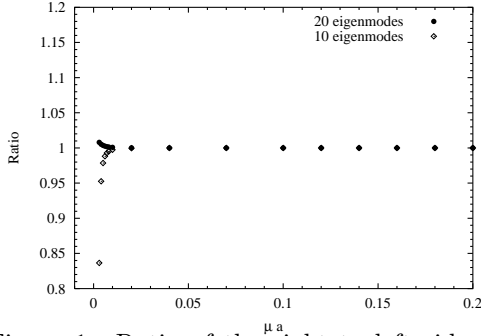}
\vspace{2.7cm}
\caption{Ratio of the right to left side of Eq. (\ref{gor}) for a
configuration with topology. $\diamond$/$\bullet$ indicates the case with 
projection of 10/20 smallest eigenmodes.}
\label{z2qn0_r}
\end{figure}

   It is shown~\cite{ehn99b} that the generalized Gell-Mann-Oakes-Renner
 (GOR) relation
\begin{equation}   \label{gor}
\mu \int d^4x \langle \pi(x) \pi(0)\rangle = 2 \langle\bar{\Psi}\Psi\rangle,
\end{equation} 
with $\pi(x)$ being the pion interpolation field,     
is satisfied for each quark mass and volume, configuration by configuration.
We utilize this relation as a check of our numerical implementation of the 
Neuberger operator. We find that for the lattices we considered
($6^3 \times 12, \beta = 5.7, 5.85; 8^3 \times 16, \beta =5.85 $, and 
$10^3 \times 20, \beta = 5.85, 6.0$) the GOR relation is satisfied very 
well (to within 1\%) all the way down to the smallest mass $\mu a = 0.0001$ 
for the $Q = 0$ sector. 
For the $Q \neq 0$ sector, the presence of zero modes demands higher
precision for the approximation of $\epsilon (H)$. For example, we
show in Fig.~\ref{z2qn0_r} the ratio of the right to left side of 
Eq.~(\ref{gor}) for
a configuration with topology on the $6^3 \times 12$ lattice at 
$\beta = 5.7$ 
as a function of the quark mass. When only 10 smallest eigenmodes are
projected, we see that the ratio deviates from one for small quark masses.
The situation is considerably improved when 20 smallest eigenmodes are
included. The situation is better than the domain-wall fermion case
when the size of the fifth dimension is limited to $L_s = 10$ to 
48~\cite{che98}.  
\vspace*{-0.6cm}

\begin{figure}[tb]
\includegraphics{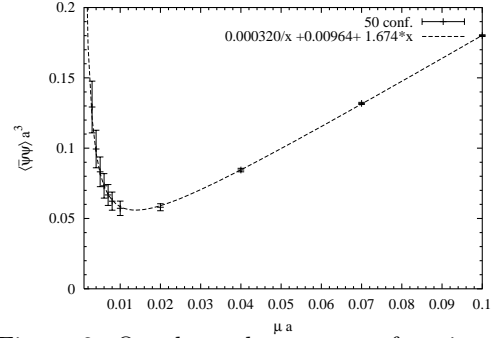}
\vspace{3.1cm}
\caption{Quark condensate as a function of the quark mass. This is done
on 50 $6^3 \times 12$ lattices at $\beta = 5.7$.}
\label{z2qw_s50}
\end{figure}

\begin{figure}[tbh]
\includegraphics{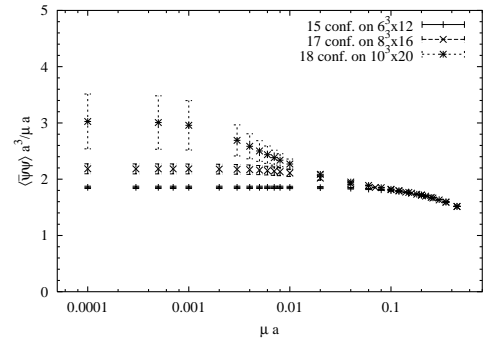}
\vspace{3.2cm}
\caption{${\langle\bar{\Psi}\Psi\rangle a^3/\mu a}$ vs $\mu a$ 
in the $Q = 0$ sector 
for 3 lattice volumes at $\beta = 5.85$.}
\label{z2qe0_u3}
\end{figure}

    We also calculate the quark condensate $\langle\bar{\Psi}\Psi\rangle$ 
with 3 to 6 $Z_2$ noises for each configuration. For small quark mass, it has
the form 
\begin{equation}  \label{con}
\langle\bar{\Psi}\Psi\rangle = \frac{\langle|Q|\rangle}{\mu V} + 
c_0 + c_1 \mu.
\end{equation}
The singular term which is due to the zero modes in the configurations
with topology ($Q \neq 0$) is specific to the quenched approximation.
It will be suppressed when the determinant is included in the dynamical
fermion case. We see this clearly in the following figure which is 
first seen with the domain-wall fermion~\cite{che98}.
A fit to the formula in Eq. (\ref{con}) is given in Fig.~\ref{z2qw_s50}. 
We see that
$c_0$ is non-zero. The standard definition of the quark chiral condensate 
entails the extrapolation of $c_0$ to the infinite volume before taking 
the massless limit. Another way is to consider the finite-size scaling
~\cite{hjl99}. When the size of the lattice is much smaller than the
pion Compton wavelength, i. e. $ L \ll 1/m_{\pi}$, the 
$\langle\bar{\Psi}\Psi\rangle$ is proportional to $\mu \Sigma^2 V$ for 
small masses besides
the $ \frac{\langle |Q|\rangle}{\mu V}$ term due to quenching. From this,
the infinite volume condensate $\Sigma$ can be extracted. We plot in
Fig.~\ref{z2qe0_u3} $\langle\bar{\Psi}\Psi\rangle a^3/\mu a$ vs 
$\mu a$ in the $Q = 0$ sector
for 3 lattice volumes ($6^3 \times 12$, $8^3 \times 16$, and 
$10^3 \times 20$) at $\beta = 5.85$. 
%\vspace*{-0.6mm}
We see that they are quite flat which indicates that
the condensate is indeed proportional to $\mu$ and we also see that they
increase with volume. 

\begin{figure}[tbh]
\includegraphics{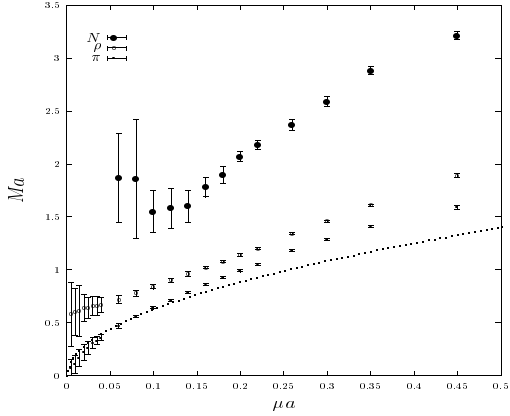}
\vspace{4.7cm}
\caption{Masses of $\pi, \rho$ and $N$ on the  $8^3 \times 16$ lattice at 
$\beta = 5.85$ are plotted vs $\mu a$.}
\label{mass_8x16}
\end{figure}
\begin{figure}[tbh]
\includegraphics{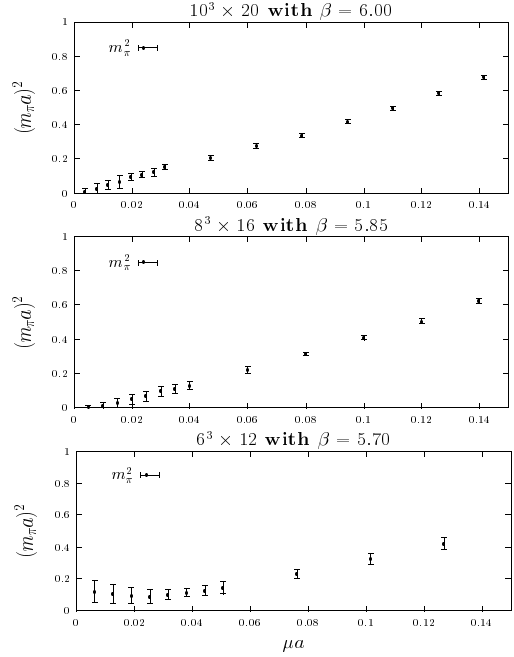}
\vspace{6.9cm}
\caption{$m_{\pi}^2 a^2$ are plotted as a function of $\mu a$ 
for three lattices
with about the same physical volume.}
\label{pi2}
\end{figure}
\begin{figure}[thb]
\includegraphics{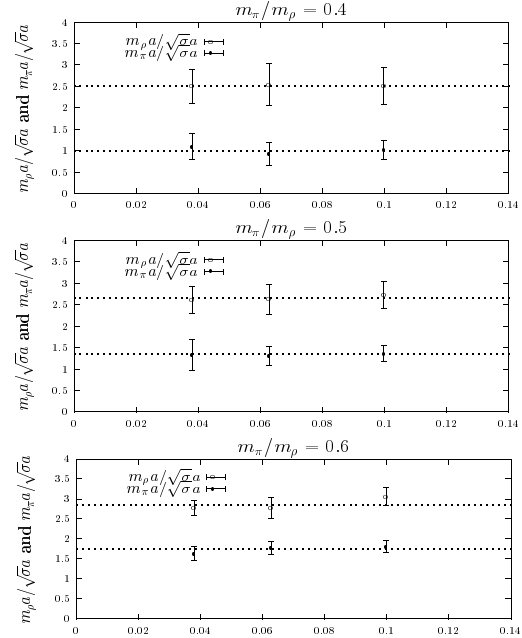}
\vspace{6.9cm}
\caption{$\pi/\sqrt{\sigma}$ and
$\rho/\sqrt{\sigma}$ are plotted vs $\sigma a^2$ for three lattices 
with similar physical volumes.}
\label{pirho_scale}
\end{figure}

%\vspace*{-1.6mm}
    We have calculated the $\pi, \rho$ and nucleon masses. A typical
result on the $8^3 \times 16$ lattice at $\beta = 5.85$ is given in
Fig.~\ref{mass_8x16}. 
We see the finite volume effect on the nucleon mass when
$\mu a$ is smaller than $\sim 0.15$.
   To see the behavior of pion masses near the chiral limit, we plot
$m_{\pi}^2 a^2$ as a function of $\mu a$ in Fig.~\ref{pi2} for three 
lattices
with about the same physical volume. It appears that there might be
a $\sqrt{\mu a}$ behavior in the very small $\mu a$ region which we will 
explore further. When we project only 10 smallest eigenmodes in the 
approximation for the sign function in the $6^3 \times 12$ case, we see that 
$m_{\pi}^2 a^2$ tends to a finite value as $\mu a \rightarrow 0$. This
implies a residual mass due to the poor approximation of $\epsilon (H)$,
a behavior similar to that observed in the domain-wall fermion with finite
$L_s$. 
   Finally, we check scaling. We plot in Fig.~\ref{pirho_scale} 
$\pi/\sqrt{\sigma}$ and
$\rho/\sqrt{\sigma}$ vs $\sigma a^2$ where $\sigma$ is the string tension
from which the lattice spacings are determined~\cite{yos98}. It is known that
the overlap operator does not have $O(a)$ artifacts. Now it appears that
the $O(a^2)$ errors are small.

This work is partially supported by 
DOE Grants DE-FG05-84ER40154 and DE-FG02-95ER40907. 
We thank R. Edwards for sharing his experience in implementing the
sign function solver. We also thank H. Neuberger for stimulating
discussions.


\vspace*{-0.1in}
\begin{thebibliography}{99}
\bibitem{lcz98} 
M. L\"{u}scher, Phys. Lett. {\bf B 428}, 342 (1998);
T. W. Chiu and S. V. Zenkin, Phys. Rev. {\bf D 59}, 074501 (1999).
\bibitem{nie99}
F. Niedermayer, Nucl. Phys. {\bf B 73}(Proc. Suppl.), 105 (1999).
\bibitem{neu98}
H. Neuberger, Phys. Lett. {\bf B 417}, 141 (1998).
\bibitem{ehn99a}
R. G. Edwards, U. M. Heller and R. Narayanan, 
Nucl. Phys. {\bf B 540}, 457 (1999).
\bibitem{ehn99b}
R. G. Edwards, U. M. Heller and R. Narayanan,
Phys. Rev. {\bf D 59}, 094510 (1999).
\bibitem{che98}
P. Chen et al., Nucl. Phys. {\bf B 73} (Proc. Suppl.), 207 (1999).
\bibitem{hjl99}
P. Hern\'{a}dez, K. Jansen, and L. Lellouch, hep-lat/9907022.
\bibitem{yos98}
T. Yoshie, Nucl. Phys. {\bf B 63}(Proc. Suppl.), 3 (1998). 
\end{thebibliography}
\end{document}